\documentclass[
superscriptaddress,
 amsmath,amssymb,
 aps,
]{revtex4-2}

\usepackage{color}

\usepackage{ulem}
\usepackage{textgreek}
\usepackage{commath}
\usepackage{gensymb}
\usepackage{float}

\usepackage{graphicx}
\usepackage{dcolumn}
\usepackage{bm}

\begin{document}

\preprint{APS/123-QED}

\title{Coherent spore dispersion via drop-leaf interaction}

\author{Zixuan Wu}
 \affiliation{Department of Mechanical and Aerospace Engineering, Cornell University, NY 14853, USA.}
\author{Saikat Basu}%
\affiliation{%
 Department of Mechanical Engineering, South Dakota State University, SD 57007, USA.
}%


\author{Seungho Kim}
\affiliation{
School of Mechanical Engineering, Pusan National University, Busan 46241, South Korea
}%

\author{Mark Sorrells}
 \affiliation{School of Integrative Plant Science, Cornell University, NY 14853, USA.}

\author{Francisco J. Beron-Vera}
 \affiliation{%
 Department of Atmospheric Sciences, University of Miami, FL 33124, USA.
}

\author{Sunghwan Jung}%
 \email{sunnyjsh@cornell.edu}
\affiliation{%
Department of Biological and Environmental Engineering, Cornell University, Ithaca, NY 14853, USA.
}%

\date{\today}


\begin{abstract}
The dispersion of plant pathogens, such as rust spores, is responsible for more than 20$\%$ of global yield loss annually, and poses a significant threat to human health. However, the release mechanics of pathogens from flexible plant surfaces into the canopy is not well understood. In this study, we investigated the interplay between leaf elasticity and raindrop momentum, revealing how it induces flow coherence and enhances spore transport with 2-10 times greater energy compared to impacts on stationary surfaces. We observed that a flexible leaf generates vortex dipoles, leading to a super-diffusive stream flow. We then developed a theoretical model that accurately predicted the average air flux from leaf edges and the vortex strength to be proportional the vibration speed of the leaves. With Lagrangian diagnostics, we further revealed the presence of hyperbolic and elliptical coherent structures around fluttering leaves, providing the dynamical description of spore transport. Our model demonstrated that a leaf aspect ratio (length/width) negatively correlates with dispersion, indicating that shorter and wider leaves promote greater pathogen spread. Additionally, we found that leaf rigidity positively correlates with dispersion due to damping effects. These mechanistic insights would help the construction of physically informed analytical models for improve local crop disease management. 
\end{abstract}


\maketitle


\section{\label{sec:level1}Introduction}
Plant pathogens (i.e., viruses, bacteria, oomycetes, and fungi) have inflicted devastating damage to fourteen major crop species that support the bulk of food production every year \cite{strange2005plant, ristaino2021persistent, skamnioti2009against,anderson2004emerging,brown2002aerial}. Specifically, biotrophic fungus species that cause commonly known rust diseases release microscopic airborne spores during the reproduction stage and execute the strategy of aerial long-distance dispersal (LDD) for intercontinental range expansion across thousands of kilometers \cite{brown2002aerial}. This airborne nature of atmospheric transport is associated with hazards that traditional plant quarantine could not resolve \cite{brown2002aerial}. From a local pathogen management perspective, more work on how environmental factors, such as raindrops, influence spore liberation can benefit understanding and stopping dispersal at its origin \cite{schmale2015highways}.

Ambient wind and rainfall have been experimentally shown to facilitate the liberation of bioaerosol through mechanical splashing and fragmentation of pathogen-bearing drops \cite{strange2005plant,schmale2015highways, hirst1963dry,huffman2013high, cevallos2012salmonella}.  Local spore transport can be achieved by wet splashing of droplets with trapped particles below 100 $\mu$m \cite{hobson1993spray}. However, larger droplets \cite{basu2017fdr} cannot sustain airborne transport from drift and have less chances of escaping the plant canopy \cite{hobson1993spray,cevallos2012dispersal,mircea2000precipitation}. 

\begin{figure*}[t]
    \begin{center}
    \centering
  \vspace{-10pt}
    \includegraphics[width=1\textwidth]{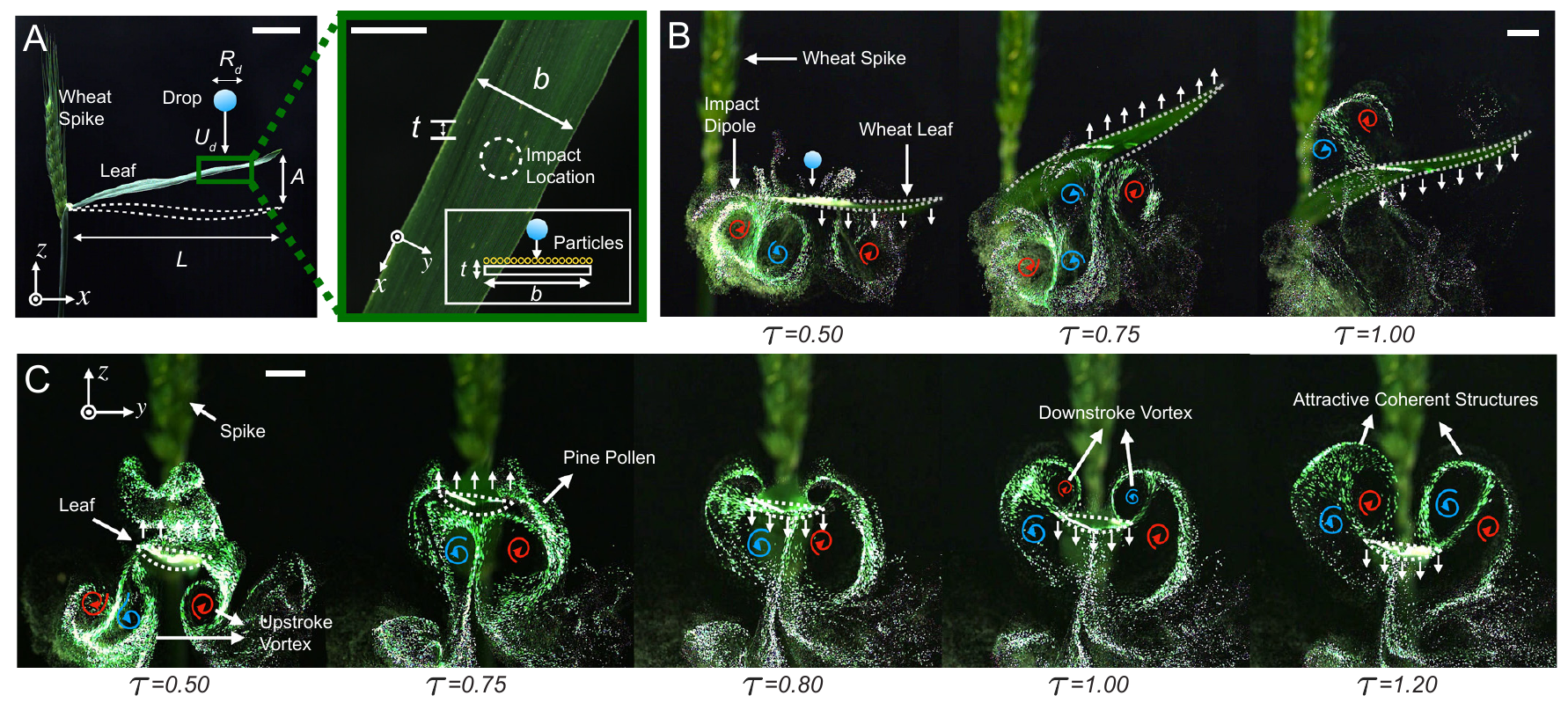}
    \caption{Vortex-induced particle dispersal on wheat leaf surfaces. \textbf{A.} Wheat leaf drop impact configurations in side view (left panel) and top view (right panel),and front view deposition schematics (inset). \textbf{B.} Side view image sequences of wheat leaf impact experiments with particle deposited, from $\tau \in [0.5\ 1.0]$ ($\tau=0$ at impact). The drop momentum and size are [$U_d$, $R_d$]=[3.13 m/s, 1.6 mm]. Color coding indicates vorticity direction. \textbf{C.} Front view image sequences of leaf-induced dispersion from $\tau\in [0.5\ 1.2]$. Corresponding videos of \textbf{B, C} are in \textcolor{blue}{SI video 1}. Scale bars are set at 50 mm for \textbf{A} right panel and 10 mm for \textbf{A} left panel and \textbf{B, C}.}
    \label{image1}
    \end{center}
\end{figure*}

Recent work on dry dispersal from raindrop-induced vortex rings shows dispersal of rust spores away from the boundary layers of a wheat leaf \cite{kim2019vortex}. However, the experiments simulate only an impact condition onto a rigid and stationary substrate. During heavy rainfall, raindrop impacts with high momentum can cause significant flapping of flexible leaves, shaped as a thin foil, generating curious flow structures regardless of the wetting conditions \cite{zhang2000flexible, gart2015droplet, orkweha2021ensemble, alben2005coherent, alben2012dynamics}. Works from the past have also shown that potential energy stored in plant structures is highly effective in bio-aerosol dispersion \cite{whitaker2010sphagnum,edwards2005record}. This leads to the question of what role drop-leaf interactions play in dispersing bio-aerosols on the surface.  

In the present work,  we studied the coupling of beam mechanics and flow dynamics to analyze the escape of spores from a vibrating leaf, triggered by raindrop impacts or ambient perturbation. We present organized particle dispersal patterns following drop impacts on leaf substrates with low flexural rigidity (10$^{-4}$--10$^{-5}$ Nm$^{2}$) \cite{niklas1999mechanical}. Wheat leaves used in this study have bending rigidity measured at $EI= 0.9\pm 0.3\times10^{-5}$ Nm$^{2}$. Cantilever vibration and field potential analysis are prescribed in the coupled vibration-vortex system on the 2D transverse plane. The mechanical details are analyzed via a parametric study with an artificial raindrop-leaf-particle system. To describe the influence of flow coherence on airborne particle transport, we apply Lagrangian diagnostics commonly adopted in geophysical transport \cite{haller2015lagrangian, serra2020search, shadden2006lagrangian} at the scale of the leaf's boundary layer, to reveal the hyperbolic and elliptical Lagrangian coherent structures (LCS) embedded in the impact-induced vortex system. Combining predictive modeling and Lagrangian metrics, we seek to reveal here the full dynamical picture that can be triggered from raindrop-leaf interactions alone, which delivers particles as parcels on ``fluid conveyor belts''. 


\section{\label{sec:level1}Results}

\subsection{\label{sec:level2} Experiments}

\begin{figure*}[t]
    \begin{center}
    \centering
  \vspace{-10pt}
    \includegraphics[width=1\textwidth]{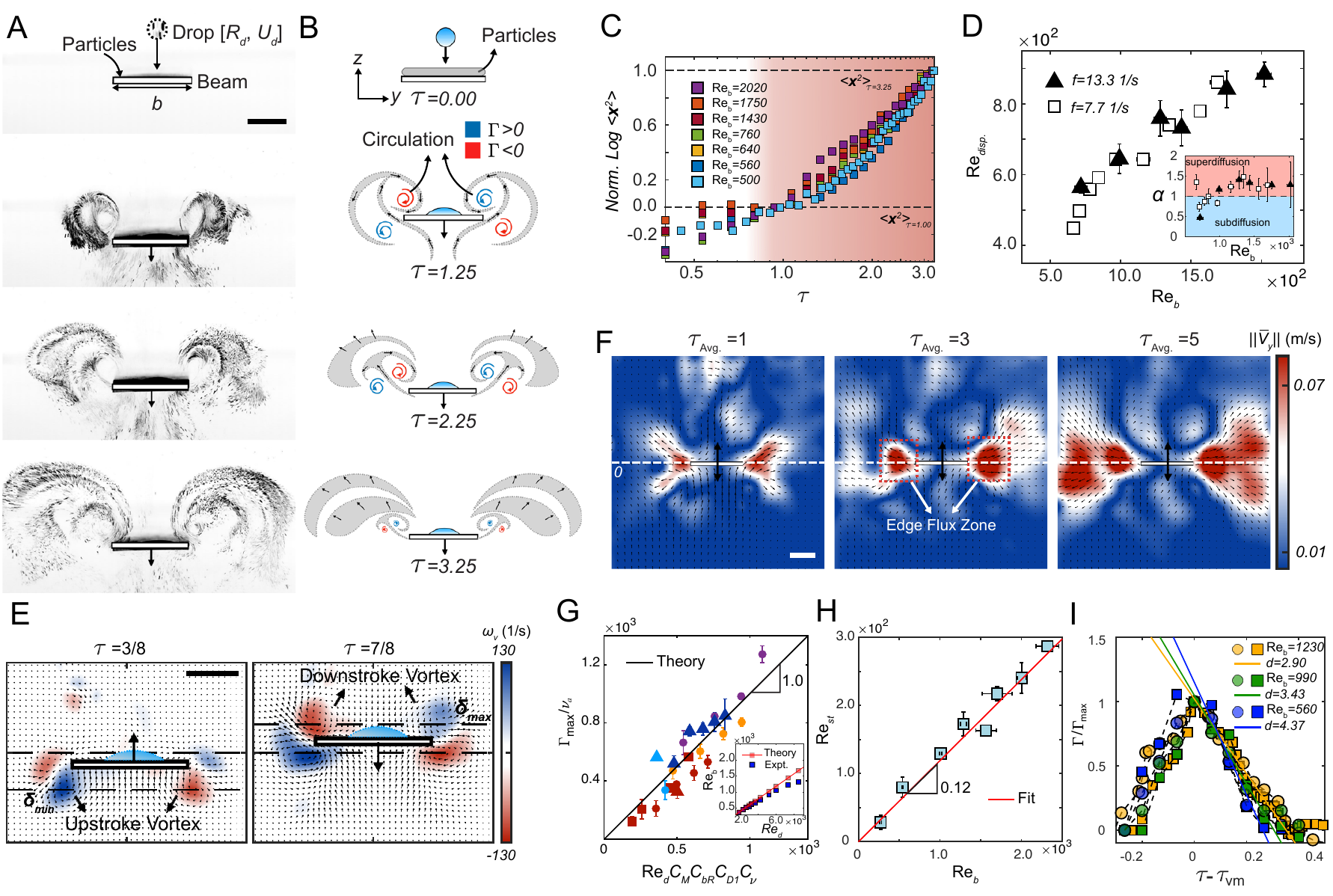}
    \caption{\textbf{A.} Flow trace visualization of dispersion on surrogate beam from $\tau=0.00-3.25$. Corresponding videos is in \textcolor{blue}{SI video 2}. Impact condition is [$U_d$, $R_d$]=[1.72 m/s, 1.60 mm], on a $L=$80 mm, $b=$20 mm beam. \textbf{B.} Corresponding schematics of dispersion steps in \textbf{A.} \textbf{C.} Normalized mean square displacement of particles from the beam center over two cycles from $\tau=0-2.25$, on log scale. It is normalized to zero at $\tau=0.75$ (beginning of active dispersion) and one at $\tau=3.25$ (the typical end of active dispersion by particles). \textbf{D.} Reynolds number of the particle dispersion across the range of $\mathrm{Re}_b$.  \textbf{E.} Vorticity field ($\omega_v$) plot of the upstroke and downstroke vortices at $\tau=3/8-7/8$.  \textbf{F.} Average horizontal velocity field $\bar{V}_y$ (in $\hat{y}$ direction) over a period of $\tau=1-5$. The velocity vector fields include the vertical direction velocity. Impact condition is [$U_d$, $R_d$]=[2.80 m/s, 1.60 mm]. \textbf{G.} Normalized circulation vs. $\mathrm{Re}_dC_{M}C_{bR}C_{D1}C_{\nu}$. Different symbols correspond to different drop-beam conditions (see Methods section). Inset here shows $\mathrm{Re}_d$ vs. $\mathrm{Re}_b$ in experiments and theory. \textbf{H.} Reynolds number of the stream vs. $\mathrm{Re}_b$. \textbf{I.} Normalized circulation (measured on left edge, normalized by the peak circulation) across time for different $\mathrm{Re}_b$ system. Time periods are aligned at the maximum circulation time $\tau_{vm}$. Scale bars are at 10 mm for all panels.  } 
    
    \label{image2}
    \end{center}
\end{figure*}


Common wheat, \textit{Triticum aestivum} (see Materials and Methods section for preparation details), is used as a representative species as it is one of the most common crops susceptible to rust infection \cite{kim2019vortex,park2020dynamics}. Wheat leaf samples are measured at width, $b= 10-20$ mm, length, $L=150-200 $ mm, and thickness, $t=0.2-0.3 $ mm (see details of wheat leaf growth and preparations in \textcolor{blue}{SI Appendix, section A}).

The drop impact experiment is conducted with a syringe pump (NE-1000, New Era Pump Systems) with DI water droplet of radius $R_d=1.2 - 2.0$ mm, released at different heights $H=0.020 - 1.20$ m onto a leaf/beam sample, as shown in Fig. \ref{image1}\textit{A}, resulting in impact velocity, $U_d=0.4-5.0$ m/s. The choice of $R_d$ and $U_d$ yields We$=\rho_d U_d^2(2R_d)/\gamma=33-1400$, which is a typical range for raindrop impacts \cite{kim2020raindrop,laws1943relation,villermaux2009single}. The longitudinal leaf axis is defined in $\hat{x}$, and the transverse leaf direction and vertical deflection are defined in $\hat{y}$ and $\hat{z}$. Side view ($xz$-plane) and top view ($xy$-plane) of the wheat sample are shown in Fig. \ref{image1}\textit{A}. A uniform, thin layer (100-200 $\mu$m) of micro-particles is deposited on substrate surface as spore surrogates, as shown in Fig.~\ref{image1}\textit{A} right panel inset.

A singular drop impact is released at 10-20 mm from the substrate tip to trigger the first-mode, free-end substrate vibration. Other impact conditions (multiple impacts, asymmetric, off-tip impacts) initiate higher vibration modes and rotations that can be approximated as a super-position of the first-mode vibration and higher modes minor in magnitudes. Asymmetric impacts empirically shed smaller, daughter vortices minor to the primary generation. Therefore, the vortex dynamics in the first-mode vibration is the basis of dispersion that is focused here. Details of the variant impact conditions are characterized and summarized in \textcolor{blue}{SI Appendix, section B}. 

Energy of the system is injected via impinging drop momentum, then converting into airflow energy via the elastic potentials of the beam. For non-dimensional analysis, Reynolds number of the drop is defined as $\mathrm{Re_d}=  U_d R_d/\nu_{d}\approx$ $10^{3}\sim$ $10^{4}$ where $\nu_{d}$ is water kinematic viscosity taken as 8.9 $\times$ $10^{-7}$ $\mathrm{m}^2\mathrm{/s}$. Reynolds number of the beam/leaf vibration is defined as $\mathrm{Re_b}=\bar{V}_b L/\nu_{a}\approx$ $10^{2}\sim$ 1.5 $\times$ $10^{3}$ where $\nu_{a}$ is the air kinematic viscosity as 1.5 $\times$ 10$^{-5}$ m$^{2}$/s. Here, the averaged beam speed is defined as $\bar{V}_b\approx 2(\delta_{max}-\delta_{min})f$, where $\delta_{max}$ and $\delta_{min}$ are the maximum and minimum deflection of the leaf substrate; $f$ is the first-mode natural frequency of vibration. Particle Stokes number is defined as $\mathrm{St}=2/9(t_{p}/t_{f})$. $t_{p}=\rho_{p}r_{p}^{2}/\mu_{a}$, is the particle relaxation time (Stokes time) where $\rho_{p}$, $r_{p}$, $\mu_{a}$ are the particle density, radius, and the dynamics viscosity of air respectively. $t_{f}= \bar{R}_{v} /\bar{V}_b$, is the characteristic time of the carrying flow where the average vortex radius $\bar{R}_{v} \approx A$ and $\bar{V}_b$ are chosen as characteristic length and velocity as the particles are dispersed via vortices.





Digital particle image velocimetry (DPIV) with smoke particles are used to extract the carrying fluid (air) velocity and vorticity fields, $\mathbf{u}(\mathbf{x},t)$ and $\mathbf{\omega_v}(\mathbf{x},t)$. Particle tracking velocimetry with glass micro-particles and pollen's is conducted to extract particle trajectories (see details in Methods section and \textcolor{blue}{SI Appendix, section A}).


\subsection{\label{sec:level2} Spore dispersion in impact-induced vibration}

For spore liberation, spores are initially hygroscopically loosened up at the mature reproductive stage, allowing further release \cite{jarvis1962dispersal}. During drop impacts, surface vortices are generated and the spreading drop collides dynamically with the spores with forces ($\sim$ 10 nN) above the inter-particle cohesion ($\sim$ 0.7 nN) \cite{kim2019vortex}, loosening spores for dispersion. Vibration generates further vortices at the two side edges and dislodges spores into surrounding vortices. The transport from leaf surface to vortices is discussed in detail in the \textcolor{blue}{SI Appendix, section C}, where three mechanisms are discussed: impact drop collision, impact vortex diffusion, and edge vortex attraction. The following analysis focuses on post-detachment delivery right after entrance into the ambient vortex flow. 

Therefore, with such mechanisms above, initial impact and the first downstroke bring particles into the boundary layers and surrounding vortices at $\tau=t/T=0-0.25$, with $\tau=0$ defined at impact. Here, $\tau$ is dimensionless time normalized by the time period $T=1/f$. At $\tau=0.25-0.50$, sudden change in acceleration leads to the shedding of the impact vortex ring along with a stroke-reversal (SR) vortex of the opposite circulation as a dipole pair. Similar vortex dynamics in flapping is documented in the literature \cite{lee2013wake,alben2005coherent,alben2008flapping}. The side view of such structure is visualized at $\tau=0.5$ in Fig.~\ref{image1}\textit{B}, with front view in Fig. \ref{image1}\textit{C} at $\tau=0.5$.
The shed vortex dipole can be seen in vorticity fields in Fig.~\ref{image2}\textit{E} at $\tau=3/8$. 
During the subsequent upstroke motion, $\tau=0.50-0.75$, another upstroke vortex is generated and follows the leaf substrate upward until $\tau=0.75$ at $\delta_{\mathrm{max}}$, the highest position of the substrate. This sequence is shown in Fig.~\ref{image1}\textit{B, C}. 

Immediately after the substrate reaches the peak, $\tau=$0.75--1.0, similar stroke reversal shedding dynamics is initiated to complete the cycle. The upstroke and downstroke vortices form a counter-rotating dipole during shedding as shown in Fig.~\ref{image1}\textit{B, C} right panels, confirmed by vorticity field in Fig.~\ref{image2}\textit{E} at $\tau=7/8$. Preferential concentration of particles at certain regions is observed to develop, as particles are transported outward. This is shown in Fig.~\ref{image1}\textit{C} at $\tau=$0.8-1.2, where particles form clustered structures as they expand outward in time. This is a clear indication of coherent flow development. 

To describe such coherent flows in the dynamics, we utilize the concept of Lagrangian coherent structure (LCS), a set of fluid parcels with attractive or repulsive properties for neighboring particles \cite{haller2015lagrangian}. The growth of these coherent profiles enhances mixing, divides up flow regions and ejects particles in specific pathways. The repetition of the described shedding cycle, enabled by leaf elasticity, produces an outward flow stream with nested layers of LCS, in which the particle cluster grows and expands under a defined dynamical sequence. Therefore, detailed LCS diagnostics is needed and used in later section to reveal the delivery pathways. 

The wake patterns under the $\mathrm{Re}_b$ tested are in a transition regime between 2S and 2P \cite{williamson1988vortex}, depending on the vibration amplitude. We primarily focus on the low-amplitude 2S cases while it should be noted that higher shedding modes exist. For the 2S scenario, flow asymmetry is observed in the shedding stream about the leaf width axis. Traditionally asymmetry is primarily induced by flow mechanics and beam geometry \cite{ebrahimi2019wake}. However, we empirically observe that asymmetry is introduced by two factors here. First, gravity deflection on drop and beam causes asymmetric vibration profile. Second, time separation of the peak vortex strength between the newer downstroke vortex and the older, decaying upstroke vortex, biases the shedding angel at 45$\degree{}$ above width axis. This lifts the center shearing layer upward as shown in Fig.~\ref{image1}\textit{C} at $\tau=$1.2. Therefore, asymmetric shedding is observed here and in later LCS analysis.

In the following analysis, we first parametrically investigate the relationship between $\mathrm{Re}_d$, $\mathrm{Re}_b$, vorticity, and dispersion efficiency of the generated flow. A reduced-order free-end vibration model is built experimentally and theoretically with thin, poly-carbonate cantilever beams to simulate the first-mode leaf vibration. Wheat leaves have high aspect ratios $C_{bL}=L/b=2-8$, which makes the thin-beam surrogate model appropriate (See discussion section for how the current modeling extends to lower $C_{bL}$ leaf systems, unlike wheat).
\begin{figure*}[t]
    \begin{center}
    \centering
  \vspace{-10pt}
    \includegraphics[width=1\textwidth]{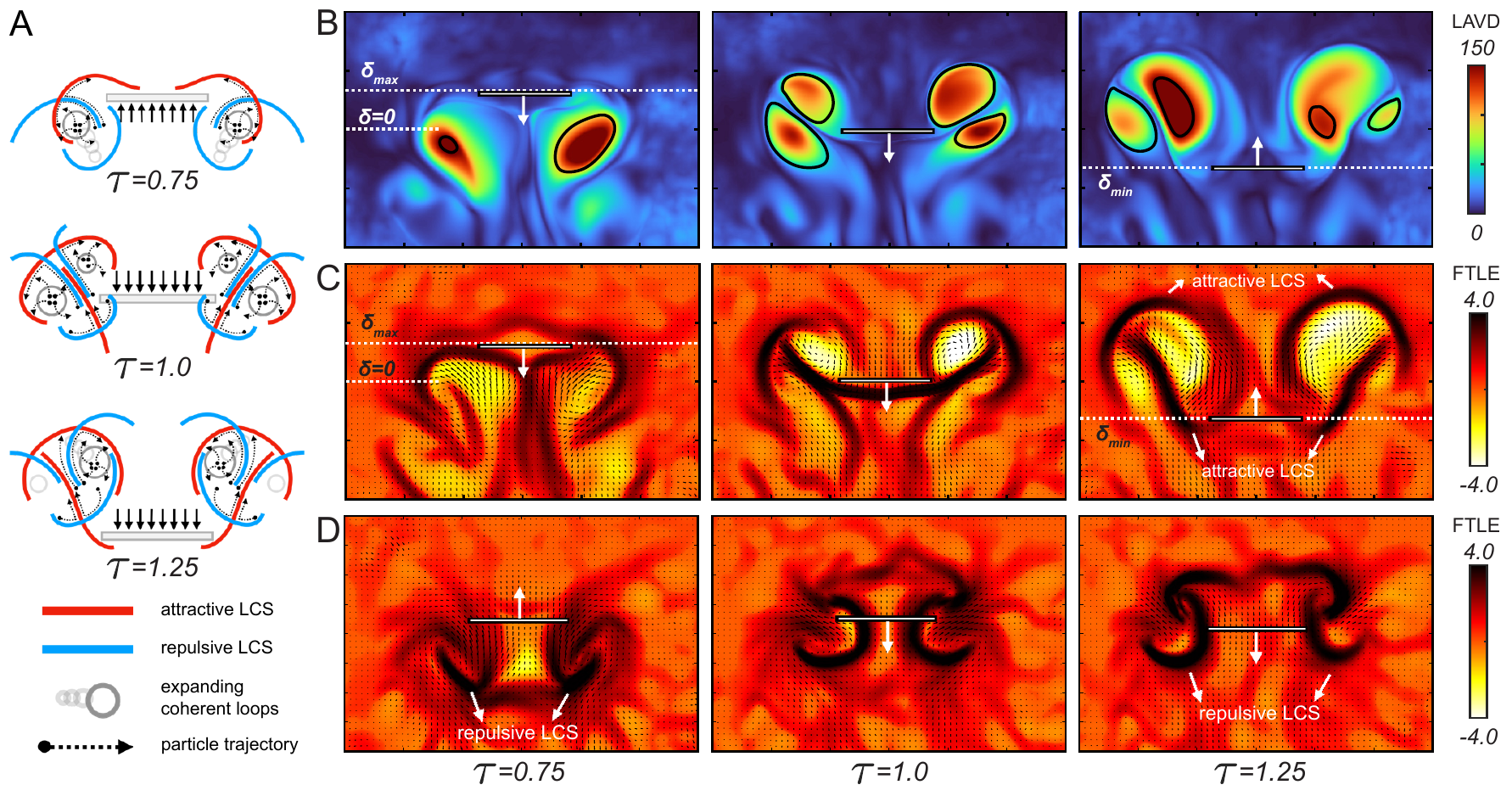}
    \caption{\textbf{A.} Schematics of the hyperbolic Lagrangian coherent structures and coherent vortex evolving over time period of $\tau=0.75-1.25$. Particle trajectories are included in the evolution. \textbf{B.} Laveraged vorticity deviation (LAVD) scalar fields (unitless) for the first full downstroke from max beam position in $+\hat{z}$ to min beam position in $-\hat{z}$, at $\tau=0.75-1.25$ (LAVD calculation details in \textcolor{blue}{SI Appendix, section E}). $\delta_{max}$ and $\delta_{min}$ labels the maximum and minimum beam location repsectively and $\delta_{0}$ labels the original beam position at $t=0$. \textbf{C.} Backward finite-time Lyapunov exponent (b-FTLE) scalar fields (unit of frame$^{-1}=3000\ \mathrm{s}^{-1}$) for the same same time sequence ( calculation details in \textcolor{blue}{SI Appendix, section F}). The attractive LCS are highlighted by high FTLE regions. Movie of the LAVD sequence in \textcolor{blue}{SI video 3}; movie of the FTLE sequence in \textcolor{blue}{SI video 4}. The analyses are conducted on the same experiment, with integration periods are both $\tau=-0.17$. \textbf{D.} Forward finite-time Lyapunov exponent (f-FTLE) scalar fields for the same same time sequence (calculation details in \textcolor{blue}{SI Appendix, section F}). The repulsive LCS are highlighted by the high FTLE regions here. The integration period is $\tau=0.17$ here.
    }
    \label{image3}
    \end{center}
\end{figure*}

\subsection{\label{sec:level2} Vortex system and dispersion capacity}
Using a beam surrogate model, dispersion stream flow from vibrating surface is visualized experimentally as shown in Fig.~\ref{image2}\textit{A} (see corresponding video in \textcolor{blue}{SI video 2}). The corresponding schematics illustrating the dispersion process are presented in Fig.~\ref{image2}\textit{B}. These figures demonstrate the observed patterns of dipole shedding, which exhibit a bias towards the upper plane, as described in the previous section. To quantitatively analyze the dispersion, we calculated the mean square displacement (MSD) $\langle \textbf{x}^2 \rangle$ of the particle clusters at different $\mathrm{Re}_b$, as shown in Fig.~\ref{image2}\textit{C}. The normalized MDS is defined as $\frac{\langle \textbf{x}^2 \rangle-\langle \textbf{x}^2 \rangle_{\tau=1.00}}{\langle \textbf{x}^2 \rangle_{\tau=3.25}-\langle \textbf{x}^2 \rangle_{\tau=1.00}}$). This highlights an active dispersion period that can be observed in Fig.~\ref{image2}\textit{A}. Details of MSD extraction can be found in \textcolor{blue}{SI Appendix, section D}. 

The relation between $\mathrm{Re}_b$ and average particle dispersion speed $\bar{V}_{disp.}$, (non-dimensionalized as $\mathrm{Re}_{disp.}=\bar{V}_{disp.}L/\nu_a$), is also extracted and presented in Fig.~\ref{image2}\textit{D}. We approximated the dispersion speed $\bar{V}_{disp.}$ as $\bar{V}_{disp.} \approx \langle \textbf{x}^2 \rangle/\tau$, the overall average dispersion rate. A monotonically increasing trend is obtained empirically with beams of different rigidity, with $\mathrm{Re}_{disp.} \approx \mathrm{Re}_b/2$ in a linear approximation. Extracting MSD in relation with time, $\langle \textbf{x}^2 \rangle \sim t^\alpha$, $\alpha$ is measured to be between 0.5-1.8, indicating a mixed diffusion and advection process for the range of $\mathrm{Re}_{b}$ tested. Above $\mathrm{Re}_{b}\approx$800-1000, as distinguished by the color coding, the dispersion flow is consistently super-diffusive. The strengthening role of advection with $\mathrm{Re}_{b}$ reassures that the beam vibration dictates dispersion kinetics.



To understand the origin of this stream flow, average velocity fields over $\tau \in [1.0\ 5.0]$ are extracted from DPIV, shown in Fig.~\ref{image2}\textit{F}, indicating increasing outward $|\hat{y}|$ velocity, $\bar{V}_y$, in the field. A cone-shaped advection corridor is observed with the average flow field, proving the existence of a vibration-generated stream flow that expands outward. We also observed edge flux zone denoted by high outward velocity near the two edges. A mechanical model is thus constructed based on 2D beam potential and drop-beam kinematics to model the average velocity magnitude of the edge flux zones with drop inertia, beam conditions, and the generated vorticity.

We define here the complex coordinates on the $yz$-plane as $\zeta=y+iz$, the complex velocity as $\chi=V_y-iV_z$ and the complex potential as $\Phi=\phi+i\psi$. By applying boundary condition $V_{z,\pm}=V_b(t)$ on the plate, where $V_b(t)$ and $V_{z,\pm}$ are vertical beam velocity and solution for $V_z$ directly above and below the beam, respectively, solution of $V_y$, $V_z$ on a thin vibrating beam is obtained as $V_{y,\pm} =\pm V_b \frac{y}
{\sqrt{(b^2-y^2)}};\  V_{z,\pm} =V_b$. 
By calculating the circulation on the left edge vortex, we obtained $\Gamma_{\mathrm{L}}=b\,V_b(t)$ (details of complex potential analysis is placed in \textcolor{blue}{SI Appendix, section D}).


We then couple it with the drop-beam interactions, with $\delta(t) \sim (U_d/f) e^{\varsigma \omega t}\mathrm{sin}(\omega t)$ \cite{gart2015droplet}, where $\omega$ and  $\varsigma$ are the $1^{\mathrm{st}}$ mode natural frequency and a damping coefficient, respectively. Evaluating at $\tau=0.5$ (maximum circulation over the damped vibrations), we obtain
\begin{equation}  \label{delta Ub}
\lvert \Gamma_{\mathrm{max}}\rvert/\nu_a=\mathrm{Re}_dC_{M}C_{bR}C_{D1}C_{\nu}.
\end{equation}

Here, $C_M=\frac{m_d}{2m_d+m_b}$, where $m_{d}$ and $m_{b}$ are the drop and beam mass, respectively; $C_{bR}=b/R_d$ is the width-drop-radius size ratio; $C_{D1}=\pi e^{-\pi \varsigma}$, a constant with damping coefficient; and $C_{\nu}=\nu_d/\nu_a$, ratio of drop-air kinematic viscosity. The theoretical derivation is corroborated by experiments, as shown in Fig.~\ref{image2}\textit{G}, in which circulations tested from different drop-beam conditions collapse onto the predictions. Reynolds number of the beam can be predicted as 
$\mathrm{Re}_{b}=\mathrm{Re}_{d}C_{M}C_{LR}C_{D2}C_{\nu}$, where $C_{LR}=L/R_d$ and $C_{D2}=e^{-3\pi/2\varsigma}+e^{-\pi/2 \varsigma}$. The relation is confirmed from the inset of Fig.~\ref{image2}\textit{G}. Therefore, we can also derive an average circulation strength over the first period as $\frac{\bar{\Gamma}}{\nu_a}=\frac{b \bar{V}_b}{\nu_a}=\mathrm{Re}_{b} \frac{C_{LR}}{C_{bL}}$.

The stream flow originates in the $v_{\theta}$ velocity component of the upstroke and downstroke vortices when they follow and shed off of the beam edge. These two counter-rotating vortices both provide an outward $|\hat{y}|$ flux on the beam edges, resulting as the edge flux zone (in red) in Fig.~\ref{image2}\textit{F}. Therefore, to model the average stream flux speed on the edge, $\bar{V}_{st}$, we assume two separated Rankine vortices and integrates the time-average $\hat{y}$ flux as the sum of time-average angular velocity $\bar{v}_{\theta}$: $2\int \bar{v}_{\theta} dr=\frac{\bar{\Gamma}}{\pi}\int_0^{\bar{R}_v} \frac{r}{\bar{R}_v^2}dr={b\bar{V}_b}/({2\pi})$, where $\bar{R}_v$ is the average radius of the circulation (see schematic for the edge flux modeling in \textcolor{blue}{SI Appendix, section D}).
This $\hat{y}$ edge flux is approximated experimentally by integrating the average $\bar{V}_y$ vertically around the edge flux zone shown in Fig.~\ref{image2}\textit{F} as $b\bar{V}_{st}=\int_{b/2}^{b/2}\bar{V}_y dl$. Integration line segment $z\in[-b/2\ b/2]$ is chosen as it covers the edge flux zone well for all cases. A ratio of the corresponding Reynolds number, $\mathrm{Re}_{st}=\bar{V}_{st}L/\nu_a$, to the beam Reynolds number becomes $\mathrm{Re}_{st}/ \mathrm{Re}_b=\bar{V}_{st}/ \bar{V}_b=1/(2\pi) \approx 0.16$. Experimentally, the slope is obtained as 0.12 (Fig.~\ref{image2}\textit{H}), a decent agreement considering variability in vortex locations. The linear relationship is corroborated by previous studies on jet stream in the longitudinal direction \cite{ebrahimi2019wake}. 

Lastly, the shed vortices show a rapid decay that can be approximated linearly, following a relation of $\Gamma / \Gamma_{\mathrm{max}}=-d(\tau-\tau_{vm})$, in which $\tau_{vm}$ is the time of peak circulation and $d$ is a dimensionless decay rate (2.5$\sim$4.5) that decreases with increasing $\mathrm{Re}_b$, as shown in Fig.~\ref{image2}\textit{I}. This is reasonable as faster stream flux reduces vortex annihilation. The vortices are created and get dissipated quickly in $\tau<0.5$. Therefore, particle dispersion is carried out by the stream flow generated, via a defined dynamical process described by LCS in the next two sections, and not by individual traveling vortices.

\subsection{\label{sec:level2} Spore expulsion by elliptical LCS}

To investigate the dynamics of dispersion, particularly the downstroke leading to active dispersion ($\tau=0.75-1.25$), two types of LCS, elliptical and hyperbolic Lagrangian coherent structures, are used. We utilized elliptical structures, or referred as rotationally coherent vorticies (RCV) \cite{haller2016lavd} below, to objectively describe the vortex structure and its role in spore expulsion. For its diagnosis, Lagrangian averaged vorticity deviation (LAVD), an objective quantity defined by 
\begin{equation}  \label{LAVD}
\mathrm{LAVD}_{t_0} ^{t_0+\tau_iT} (\mathbf x_0) = \int_{t_0} ^{t_0+\tau_iT} |\Omega(F_{t_0}^{t}(\mathbf x_0), t)-\bar\Omega(t)|\,dt
\end{equation} is used. The method objectively identifies the vortices in the unsteady flow, by finding the concentrated
high-vorticity regions from integration of $t_0$ to $t_0+\tau_iT$, with $t_0$ as the start of integration, and $\tau_i$ the dimensionless integration period. $\Omega(F^t_{t_0}(\mathbf x_0),t)$ denotes the vorticity of the fluid over the flow map $F^t_{t_0}$, and $\bar{\Omega}$ is the vorticity at time $t$ averaged over the tracked fluid bulk. 



Empirically, $|\tau_i| =0.15 \sim 0.25$ is the integration time that captures the fluid structures in a cycle, as vortex growth and shedding occur within a quarter stroke $\tau=1/4$. The resulting LAVD map is shown in Fig. \ref{image3}\textit{B} for the first downstroke $\tau=0.75-1.25$ (see movie in \textcolor{blue}{SI movie S3}). Boundaries of the coherent vortices are calculated and marked in Fig. \ref{image3}\textit{B} (black outlines). 

\begin{figure*}[t]
    \begin{center}
    \centering
  \vspace{-10pt}
    \includegraphics[width=1\textwidth]{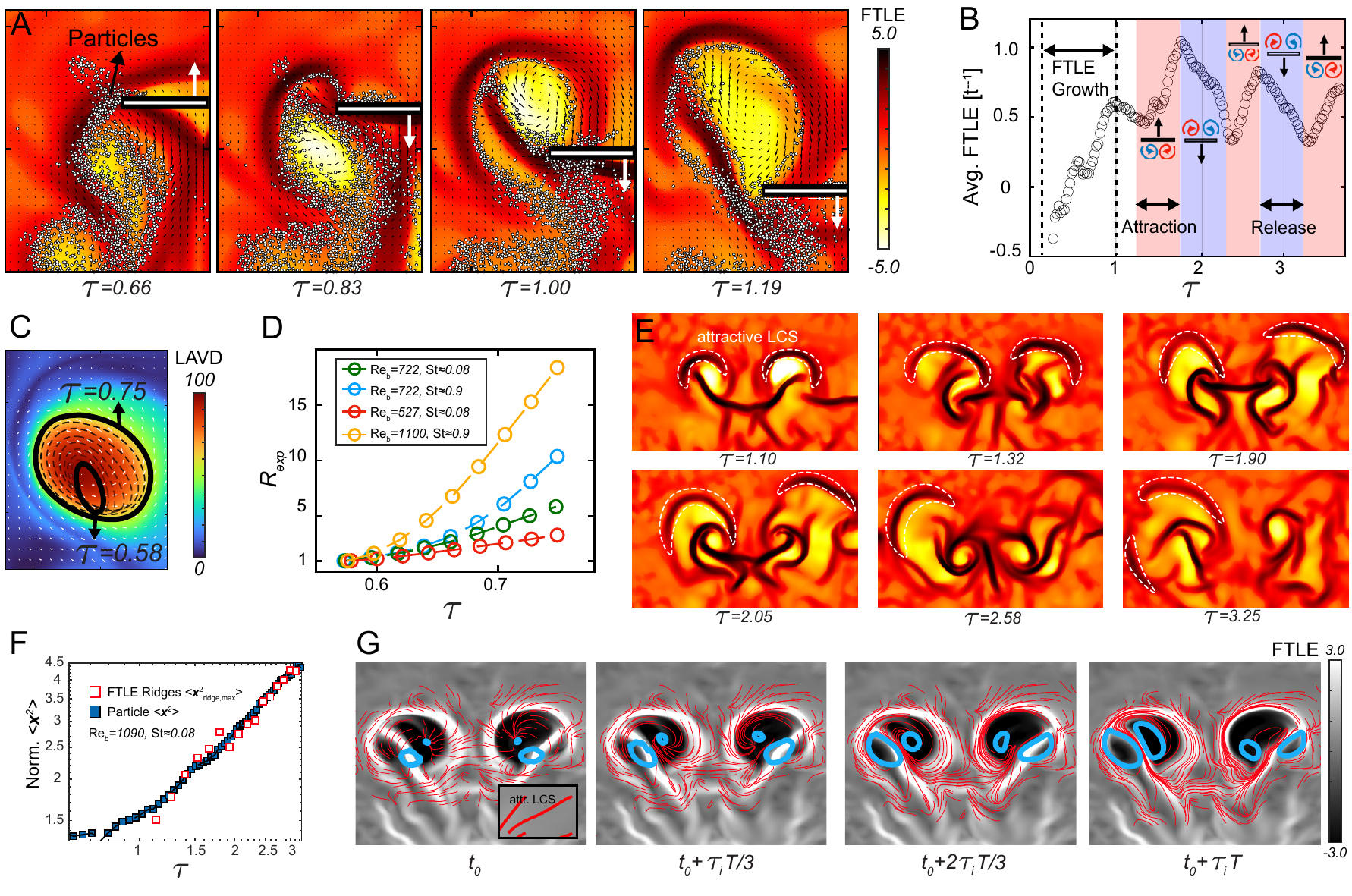}
    \caption{\textbf{A.} Experimental particle (pollen) tracking overlaid on background FTLE fields for $\tau=0.66-1.19$, [$U_d$, $R_d$]=[2.97 ms$^{-1}$, 1.60 mm]. 
    \textbf{B.} Average FTLE (unit of frame$^{-1}=3000\ \mathrm{s}^{-1}$) of particles for $\tau=0.0-4.0$, [$U_d$, $R_d$]=[2.97 ms$^{-1}$, 1.60 mm]. The LCS growth period from $\tau=0-1$ and the upstroke particle attraction, downstroke particle release periods are labeled. \textbf{C.} Experimental (pollen) time-series of coherent vortex expansion for $\tau=0.57-0.75$; the background LAVD is obtained at $\tau=0.75$ (end of upstroke dispersion) and an integration period of $\tau=-0.17$. \textbf{D.} Coherent vortex (RCV) expansion for $\tau\in [0.57\ 0.75]$, at different particle Stokes number and drop Reynolds number. \textbf{E.} Advection of the attractive LCS at $\tau=1.10-3.25$. \textbf{F.} Normalized MSD for particles and FTLE ridge locations. Both are normalized by the beam width $b^2$. Same video source as in \textbf{E}. Reynolds number of the beam and Stokes number of particle is listed. \textbf{G.} Migration of actual attractive LCS (red lines) onto high FTLE regions (white in background), and the migration of coherent vortices (in blue) over $\tau=\in [1.08\ 1.34]$ (downstroke). The high FTLE regions corresponds to that of Fig. $\ref{image3}$.}
    \label{image4}
    \end{center}
\end{figure*}

In order to characterize the expulsion flux from such vortex, we define a flux criterion $\mathcal{F}$ to describe the inertial particle ejections in Eq. \ref{flux} for the fluid regions within said boundary, labeled as $\mathcal{V} (t)$. (see full LAVD sequence and details of LAVD, flux calculations in \textcolor{blue}{SI Appendix, section E}). \begin{equation}  \label{flux}
\mathcal{F} \propto t_p \frac{1-R_{\rho}}{1+R_{\rho}/2} \int _{\mathcal{V}(t)} QdS.
\end{equation} 
Q is the Okubo-Weiss criterion \cite{provenzale1999transport}, defined here as $\omega_v^2-S_s^2-S_n^2$. $\omega_v$ is the relative vorticity, $S_s$ is the shear strain, and $S_n$ is the normal strain. Inside a Lagrangian vortex, One can expect $Q>$0 \cite{provenzale1999transport}. The flux calculation thus predicts a positive outward flux $\mathcal{F}$ from vortex centers for inertial particles with density ratio $R_{\rho}=\rho_{a}/\rho_{p}\ll 1$, where $\rho_a$ is the air density. Indeed, as traditional coherent vortices tend to trap ideal tracers and do not expand, experiments with inertial particles here demonstrate strong particle dispersion behavior as the coherent vortex boundary expands, shown in Fig.~\ref{image4}\textit{C} for $\tau=0.57-0.75$. Therefore, the coherent vortices identified in Fig. \ref{image3} effectively serve as traveling sources of outward flux for spores around leaves.

 
Strength of flux increases in proportion to the particle response time $t_p\propto t_f St $, effectively the Stokes number, and $\int QdS$. Such flux relation is experimentally validated in Fig.~\ref{image4}\textit{D} by the boundary expansion ratio of the coherent vortex, $R_{exp}=S/S_{0}$, in which $S_0$ is the coherent vortex size before expansion and $S$ is the expanded size at $\tau=$0.75. Sample system with higher $\mathrm{Re}_b$ and St, denoting stronger circulation and particle inertia, display the highest expansion on average as shown.

\subsection{\label{sec:level2} Spore transport by hyperbolic LCS}

Interactions of vortex structures during their growth and shedding organize the airflow near the leaf into nested hyperbolic LCS, which attract or repel particles readily ejected by the vortices. 

To identify hyperbolic LCS, finite-time Lyapunov exponent (FTLE) diagnostics is initially applied, outputting the flow separation rate for the 2D $yz$ domain (see calculation details in \textcolor{blue}{SI Appendix, section F}). In brief, the calculation takes a infinitesimal perturbation around a point $\textbf{x}(t_0)$, expressed as $| \delta\mathbf x (t_0) |$, and extract the exponent of the perturbation growth $\sigma$ in time $\tau_iT$:

\begin{equation} \label{ftle}
|\delta\mathbf x (t_0+\tau_iT) |=e^{\sigma \tau_i T}| \delta\mathbf x (t_0) |.
\end{equation}

Observing the dynamics backward in time $\tau_i<0$, regions with the largest perturbation growth (high $\sigma$) reveal the most attractive surfaces as they pull together fluid elements furthest apart, namely, the attractive LCS. They primarily concentrate in the red regions (high FTLE lines) in Fig.~\ref{image3}\textit{C} for the downstroke. Equivalently, regions with maximum repulsion, i.e. the repulsive LCS, is obtained with forward integration $\tau_i>0$, and reside mainly in the red regions in Fig.~\ref{image3}\textit{D}, where particles are stretched apart the most in future time (see full sequence in \textcolor{blue}{SI Appendix, section F} and \textcolor{blue}{SI movie S4}). We will refer to such dark red regions with high FTLE values as FTLE ridges below, which typically coincide with LCS locations.

Combining with the coherent vortices, a more complete picture of spore dispersion from fluttering leaves can be depicted. Schematics in Fig.~\ref{image3}\textit{A} illustrates these dynamics under the two types of LCS. Immediately before downstroke at $\tau\approx0.75$, upstroke vortices continue to eject particles outward, indicated also by the surrounding repulsive LCS in blue curves. As the downstroke vortex grows in strength, significant shearing between the two sets of vortices develops, and their coupling creates attractive LCS that pulls particles outward at an angle as mentioned in previous discussion of Fig.~\ref{image1}. A cap-like attractive LCS then develops on the dipole exterior. It has multiple repulsive LCS penetrating through, cooperatively pulling particles outward. The process completes as the substrate reaches minimum position, and the nested hyperbolic LCS expands in size before weakening. During the downstroke, coherent vortices remain active in flux, ejecting more particles onto the nest. 

Dynamic flow attraction on particles is further validated by overlaying particle locations onto the FTLE map (backward integrated), shown in Fig.~\ref{image4}\textit{A}. Extracting the FTLE values of these particles overtime reveals the continuous development of flow coherence during the first cycle, shown in Fig.~\ref{image4}\textit{B}. FTLE increment indicates particles exiting high-vorticity regions and entering high-strain regions, resulting in particle entrapment on LCS. Cyclic rise and fall indicate that particles are pulled into attractive profiles during upstroke, and released into the surrounding in a advection-assisted diffusion process during downstroke, confirming our finding in Fig.~\ref{image2}. Therefore, particle entrapment is only momentary, as the particles are released in each cycle. 

The hyperbolic LCS can have prolonged influences on more distanced particles, since they can have a long lifetime, $\tau>3$, as they expand and travel outward, demonstrated in Fig.~\ref{image4}\textit{E}. The speed of advection is identical to that of the particle cluster boundary. This is shown in Fig.~\ref{image4}\textit{F}, where the normalized MSD over time is plotted along the outermost FTLE ridge positions, $\textbf{x}_{\mathrm{ridge,max}}^2$. This makes sense since the frontier of the super-diffusive stream flow discussed in Fig.~\ref{image2} must be an expanding, attractive LCS that pulls materials outward.  

Backward FTLE ridges for the wheat leaf samples are also extracted and shown in \textcolor{blue}{SI Appendix, section F}. Similar attractive LCS profiles to Fig.~\ref{image3}\textit{B} are displayed, validating the surrogate beam model. 


\subsection{\label{sec:level2} Flow dynamics from geodesic transport theory}
While FTLE diagnostics render the approximate locations of the LCS, ridges are merely coherence imprints that is left behind by true LCS in the flow. For a more rigorous identification, we turned to the geodesic transport theory \cite{haller2012geodesic} to calculate the attractive LCS as material lines (a set of fluid elements) in the 2D domain. For each fluid patch that is shown in the domain of Fig.~\ref{image3}\textit{B-D}, an attractive LCS (red line) can be calculated over $t \in [t_0,  t_0+\tau_i T]$ as shown in the inset of Fig.~\ref{image4}\textit{G}. In forward time, these LCS first attract particles in the fluid patch, then pulls the whole fluid patch forward with itself as the center backbone. Eventually, at the end of integration, many of them land near the FTLE ridges, an imprint left by this migration. The dynamics sequence is shown for the downstroke in Fig.~\ref{image4}\textit{G} with backward FTLE map and coherent vortices overlaid. Similar analysis is documented in literature for geophysical flows \cite{beron2015flow}. 

The landing proximity of LCS to FTLE ridges depends on the Stokes number. Particles with higher Stokes number exhibit more preferential concentration and pattern formation near the FTLE ridges, as the inertia of particles introduces bias in trajectories towards low-vorticity, high-strain regions, commonly observed in literature \cite{goto2008sweep,sapsis2010clustering}. St, $r_{p}$, and $\rho_{p}$ of common bio-aerosols and experimented particles used are reported in Table~\ref{table:st} under Methods section.




\section{\label{sec:level1}Discussion}
In this work, we studied the dynamics of how pathogenic spores escape vibrating leaf boundaries in generated flows. We discovered that leaf elasticity enabled a vortex-induced stream flow that organizes and promotes spore dispersion. With surrogate beam as a model system, we first use theory and empirical evidence to predictively model dispersion strength with drop-leaf interactions. Utilizing LCS diagnostics, we revealed the full dynamical pathways embedded in the generated stream transport by identifying the nested attraction and repulsion regions. The modeling, along with the dynamics picture, proposes a mechanical explanation for the co-occurrence of rainfall and bio-aerosol dispersion in air \cite{huffman2013high}. This is directly proven as Reynolds number of the leaf vibration linearly scales with Reynolds number of the drop, $\mathrm{Re}_b\propto \mathrm{Re}_d C_M$. And Reynolds number of leaf vibration is the main tuning parameter that linearly couples with Reynolds number of the average stream flux $\mathrm{Re}_{st} \approx 0.12 \mathrm{Re}_b$, and the circulation strength $\Gamma/\nu_a \propto \mathrm{Re}_b$. Most importantly, we show that the average particle dispersion speed scales with beam speed as $\mathrm{Re}_{disp.}\sim \mathrm{Re}_{b}/2$ under linear approximation. We thus observe empirically dispersion rate increases with $\mathrm{Re}_d$. Therefore, our work directly provides physical explanation for more effective bio-aerosols dispersion during rainfall as observed. 







While leaf elasticity enables the flow generation, vibration frequency $f$ itself do not directly influence leaf vibration speed $\mathrm{Re}_b$. A simple scaling shows that amplitude and frequency are coupled as $A\sim U_d/f$, based on the balance of elastic potential and input drop energy as $E_d=m_dU_d^2/2=E_b\sim m_b(Af)^2$ (see details of derivation in \cite{gart2015droplet}). And thus $V_b\sim Af \sim U_df/f\sim U_d$, a function of $U_d$ and prefactors only. Such result holds for drop impacts and ambient perturbations as long as the initial elastic deformation is the sole source of potential energy input for the substrate, without consideration for damping. Therefore, leaves of different natural frequency $f$, can similarly vibrate and disperse under this mechanism, and the Reynolds numbers $\mathrm{Re}_{b}$ and $\mathrm{Re}_{disp.}$ only scale with $\mathrm{Re}_d$ for early cycles $\tau<$1.

The latter cycles $\tau>$1 however, is strongly affected by the damping coefficient, $\varsigma$, as energy loss becomes significant. A significant portion of the energy is dissipated in generating airflow, thus air damping is three orders of magnitude higher than other damping factors \cite{gart2015droplet}. $\varsigma$ scales with size and frequency parameters as $\varsigma\sim \frac{\mu_a}{\rho_b t bf}$ \cite{gart2015droplet}. Therefore, air damping effect is inversely related to vibration frequency. Since we know that $f \propto \sqrt{\frac{EI}{m_b}}\frac{1}{L^{1.5}} = \sqrt{\frac{EI}{\rho_b t b}}\frac{1}{L^2}$ ($\rho_b$ is the beam density and $t$ is the substrate thickness) \cite{gart2015droplet}, a leaf structure with a higher flexural rigidity $EI$ thus provides more vibration-based dispersion beyond the first cycle. 

Damping also comes into play as the leaf shape varies. The aspect ratio, $C_{bL}=L/b$, is an important shape parameter of the reduced-order leaf model. Assuming both area and mass of the leaf are constant, the beam speed, $V_b$, is independent of $C_{bL}$ without damping. By considering air damping, however, scaling analysis shows that $\omega \sim L^{-2}$ and $b \sim L^{-1}$, and thus $\varsigma \sim C_{bL}^{3/2}$ in the first order. It shows that the averaged beam speed, $\bar{V}_b$, decreases with an increasing aspect ratio by considering the air damping. Conversely, a beam with a low aspect ratio promotes the dispersion of particles in the air. However, these particles need to travel across the leaf width in order to become suspended (see details in \textcolor{blue}{SI Appendix, section C}). A critical width scale thus exists as $b_c \sim R_d \mathrm{We}^{1/2}$, for direct collision ejection of particles off the surface.

We account for three dimensional effects here. First, we observe empirically that substrate velocity and thus vorticity decreases as we move away from leaf tip in the $\hat{x}$ direction towards the stem. Theoretically, drop impacts at location of $x$ measured from leaf stem, yield vibration velocity of $\frac{V_{b, x}}{V_{b, tip}}=(3k^2-k^3)/2$, with ratio $k$ defined as $k=x/L$, as discussed in \textcolor{blue}{SI Appendix, section B}. Therefore, impacts away from the tip simply diminish vibration and dispersion speed. Second, background flow in $\hat{y}$ is observed to suppress the dispersion stream on the opposing edge, which has a flow stream going against the background velocity, but this ambient flow in $\hat{y}$ augments the stream on the other edge simultaneously. Lastly, since leaf tips at $x=L$ are typically pointy with no realistic front edge, the scenario of front edge dispersion in $\hat{x}$ is not considered here.


In summary, we have developed a comprehensive model that can accurately predict the influence of leaf properties, particle characteristics, and raindrop conditions on spore dispersion in a quiescent environment, i.e. minimal background flow and large Strouhal number, Str$=fA/\Bar{U}_{bkg.}\gg 1$. Local acceleration of the vibrating substrate here dominates the background advection, which allows us to parametrically analyze the drop-leaf mechanics alone. We are able to prove that without significant turbulence, the drop impact alone channels enough energy to power a stream flow as $\mathrm{Re}_{disp.}\sim \mathrm{Re}_b/2$, which becomes super-diffusive above $\mathrm{Re}_b>$1000. To further complete the energy analysis, we approximated the kinetic energy of the vortical flow as $E_\Gamma \sim \rho_a\Gamma^2L_{\Gamma}$ \cite{kim2015spontaneous}, where $L_{\Gamma}$ is the length of the connected vortex tube. We can then compare the energy budget spent in airflow generation via a drop impact on stationary vs. flexible surfaces. We could first estimate $\Gamma_{stat.}\approx 0.1 U_dR_d\mathrm{Re}_d^{3/8}\mathrm{Re}_a^{-1/4}$ (where $\mathrm{Re}_a=U_d(2R_d)/\nu_a$) from literature \cite{kim2019vortex} and obtained the expression for $\Gamma_{flex.}$ from $\Gamma_{max}$ above. Then, at varying input drop energy $E_d=$0-0.01 J (from the natural range of [$U_d$, $R_d$] in this present study), we calculated that a ratio of the rotational energy to the drop kinetic energy, $E_\Gamma/E_d$, for a flexible leaf surface ranges from 2.5-5.5$\%$, whereas for a stationary leaf the ratio is around 0.5-1.5$\%$ across the parameter space. Leaf elasticity has permitted more energy budget for dispersive vortex generation, a key role that has been largely omitted in leaf-spore dispersion mechanics. 

Therefore, the current study renews the current understanding of spore dispersal avenues \cite{dressaire2016mushrooms, gregory1949operation, roper2010dispersal, ingold1971t, kim2019vortex} and connects impact mechanics to Lagrangian coherence, uncovering an active spore dispersion mechanism that shows less reliance on passive environmental carriers such as traveling splashed droplets or background canopy currents. The work establishes the coupling of leaf elasticity and rainfall in generating a stream flow that disperses surface-bound spores upward ($+\hat{z}$) and sideways (+ or -$\hat{y}$), with defined dynamics pathways hidden within.

 \setlength{\tabcolsep}{5pt}
\renewcommand{\arraystretch}{1.2}
\begin{table*}
\caption{Physical properties of particles and Stokes number. $t_f=0.01-0.05$. Pollen and spore density are from literature \cite{roper2010dispersal,sosnoskie2009pollen}.}

    \begin{center}
        \small
        \setlength\tabcolsep{2pt}
    \begin{tabular}{ |c|c|c|c| }
    \hline
        {Particle Types (*: experimented)} &{$\rho_{p}$ (g/cm$^{3}$)} &{$r_{p}$ ($\mu$m)} & St \\[0pt]
         \hline
    
       {*Soda lime glass spheres (SGS)} & 2.5 & 10  &0.02-0.08\\
              \hline
       {*Glycerine-water smoke} & 1.0 & 1-2 &1e-4-4e-3\\
       \hline
       {*Pine (\textit{P. contorta}) pollen}& 1.2 & 20-25 & 0.1-0.9 \\
        \hline
       {Forget-Me-Not (\textit{M. palustris}) pollen} & 1.2 & 2.5-5.0 \cite{knight2010relationship} & 0.002-0.005\\
               \hline
 
       {Wheat rust (\textit{P. triticina}) spore} & 1.0  &  10 \cite{kim2019vortex}  & 0.02-0.1\\
       \hline
              \end{tabular}
           \end{center}

\label{table:st}
\end{table*}

\section{\label{sec:level1} Methods}

\subsection{\label{sec:level2} Drop-impact experiments}
High-speed photography (FASTCAM, Photron) at 1,000--3,000 fps is used. A flapping wheat leaf is mechanically modeled as an angularly flapping thin cantilever beam; thin polycarbonate beams, $\rho_b$=1,220 kg/m$^3$, are used exprimentally, whose dimensions, rigidity, and wetting conditions are documented in \textcolor{blue}{SI Appendix, section A}), along with that of the wheat samples. Mechanically, the beam substrate is fixed by clamping on one end along the longitudinal axis. The rotational degree of freedom around the longitudinal axis is thus limited with the high $L/b$ ratio used and the center impact along this axis. After securing the beam, drop impacts are induced with a syringe pump at a pumping rate of 0.2 mm/min. An impact is induced near the tip of the beam (8-10 mm) to observe the maximum impact consequences. The location is chosen to prevent significant spillage as well, since the maximum spreading radius $R_m$ can be calculated as 6 -- 11 mm, with $R_m\approx (1/2)R_d \mathrm{Re}_d^{1/4}$ from the aforementioned [$U_d$, $R_d$] conditions \cite{kim2019vortex}. Combinations of beam, drop, and impact velocity tested in Fig.~\ref{image2}G are listed in \textcolor{blue}{SI Appendix, section D}.

\subsection{\label{sec:level2} Visualization methodology}
Particle visualization employs the use of glass particles and pine pollens. They are uniformly deposited on top of the substrate surfaces prior to the drop impact experiment, with size, density, and Stokes number, St $=2/9(t_{p}/t_{f})$, reported in Table \ref{table:st}. Particle layer thickness is consistent with experimentation methodology in \cite{kim2019vortex}, at 0.1-0.2 mm. Particle size and density ranges are typically $r_p=1.0-20.0\ \mu$m and $\rho_{p}=1.0-2.5\ \times 10^3$ kg/m$^3$ respectively. 
 
 Smoke visualization is used to perform two-dimensional digital particle image velocimetry (DPIV) on the 2D transverse cross section, in order to extract the velocity and vorticity fields at the location of impact and shedding. Chauvet smoke machine is paired with a 40-60 glycerol-water mixture to produce a thick smoke layer that fills the field of view. A laser beam (sheet laser) with the intensity of 5 mW is used to illuminate the smoke layer at the 2D transverse cross-section of impact point, with laser sheet thickness of 0.2 -- 0.5 mm. DPIV is conducted with the MATLAB package PIVLab by Thielicke \cite{thielicke2014pivlab}. CLAHE is enabled with window size 64 as the only image setting. The analysis uses an FFT window deformation with 3 passes; pass 1 is integration area of 120 pixel, and 64-pixel step; pass 2 is integration area of 64 pixel, and 32 pixel-step; pass 3 is integration area of 32 pixel, and 16 pixel-step. Gauss 2X3-point estimator is used with high correlation robustness. The error of the velocity vectors are estimated to be 0.0128 m/s from difference of actual tracer measurements and DPIV analysis, an error rate of 2-10 $\%$.

\begin{acknowledgments}

Wheat samples are grown at the Plant Breeding and Genetics Section, School of Integrative Plant Breeding and Genetics Section at Cornell University. Please see wheat preparation details in the \textcolor{blue}{SI Appendix, section A}. This work was supported by the National Science Foundation Grant No. ISO-2120739. The collaboration between F.J.B.-V. and S.J. was initiated at the Aspen Center for Physics, which is supported by the National Science Foundation grant PHY-1607611.
\end{acknowledgments}




\bibliography{apssamp}

\end{document}